\documentclass{aa}  
\usepackage{graphicx}
\usepackage{txfonts}
\begin{document}
   \title{The origin of the $<\mu_e>$-$M_B$ and Kormendy relations in dwarf
   elliptical galaxies}


   \author{A. Boselli\inst{1}
           \and
           S. Boissier\inst{1}
	   \and
	   L. Cortese\inst{2}
	   \and
	   G. Gavazzi\inst{3}
          }

   \offprints{A. Boselli}

   \institute{Laboratoire d'Astrophysique de Marseille, UMR 6110 CNRS, 38 rue F. Joliot-Curie, F-13388 Marseille France\\
              \email{Alessandro.Boselli@oamp.fr; Samuel.Boissier@oamp.fr}
         \and
             School of Physics and Astronomy, Cardiff University, 5, The Parade, Cardiff CF24 3YB, UK\\
             \email{Luca.Cortese@astro.cf.ac.uk}
	 \and
	     Universita degli Studi di Milano-Bicocca, Piazza delle Scienze 3,
	     20126 Milano, Italy
	     \email{Giuseppe.Gavazzi@mib.infn.it}
             }

   \date{}

 
  \abstract
  {}  
    {The present work is aimed at studying the distribution of galaxies of different types
   and luminosities along different structural scaling relations to see whether massive 
   and dwarf ellipticals have been shaped by the same formation process.}
   {This exercise is here done by comparing the distribution of Virgo cluster massive and dwarf ellipticals and star forming
   galaxies along the B band effective surface brightness and effective radius vs. absolute magnitude relations 
   and the Kormendy relation to the predictions of models tracing the effects of ram-pressure stripping
   on disc galaxies entering the cluster environment and galaxy harassment.}
    {Dwarf ellipticals might have been formed from low luminosity, late-type spirals that recently entered into 
   the cluster and lost their gas because of a ram-pressure stripping event, stopping their activity of star
   formation. The perturbations induced by the abrupt decrease of the star formation activity are sufficient
   to modify the structural properties of disc galaxies into those of dwarf ellipticals.  
   Galaxy harassment induce a truncation of the disc and generally an increase of the effective 
   surface brightness of the perturbed galaxies. The lack of dynamical simulations of perturbed galaxies
   spanning a wide range in luminosity prevents us to drive any firm conclusion on a possible harassment-induced
   origin of the low surface brightness dwarf elliptical galaxy population
   inhabiting the Virgo cluster.}
{Although the observed scaling relations are consistent with the idea that the distribution of elliptical galaxies along the mentioned
  scaling relation is just due to a gradual variation with luminosity of the Sersic index $n$, 
  the comparison with models indicates that dwarf ellipticals might have been formed by a totally different
  process than giant ellipticals.
  }


   \authorrunning{Boselli et al.}
   \titlerunning{Scaling relations in dwarf ellipticals}
   \maketitle

\section{Introduction}

Scaling relations are often used as major constraints for models of galaxy formation and evolution.
They can be used to trace the contribution of the different stellar components 
to the total luminosity of galaxies, as in the case of the color magnitude relation 
(Visvanathan \& Sandage 1977; Bower et al. 1992 for early-type galaxies,
Tully et al. 1982, Gavazzi et al. 1996 for late-type galaxies) or to study the relationship
between kinematical, structural and stellar population properties of galaxies as in the case of 
the Tully-Fisher relation for spirals (Tully \& Fisher 1977) and the fundamental plane for ellipticals 
(Dressler et al. 1987; Djorgovski \& Davis 1987). Scaling relations have been used also to compare the 
total gas (Boselli et al. 2002), the present or past star formation activity (Boselli et al. 2001)
and the metallicity (Zaritsky et al. 1994; Bender et al. 1993) 
of galaxies with their luminosity or mass.
The study of these different scaling relations has been crucial for showing the role of mass 
in the formation of galaxies (Gavazzi et al. 1996; Boselli et al. 2001), a result now generally 
called downsizing effect, which is a new major constraint for hierarchical models of 
galaxy evolution (De Lucia et al. 2006).\\
The surface brightness vs. absolute magnitude and the Kormendy relation (effective surface 
brightness vs. effective radius; Kormendy 1985) have been often used in the literature to
study the process governing the formation of early-type galaxies. Although limited to
structural properties, these two relations can be easily determined for large samples of galaxies
including both giant and dwarf ellipticals. High resolution spectroscopic data, necessary for measuring
velocity dispersions needed in the construction of the fundamental plane, are still relatively rare 
for low luminosity systems. \\
The study of the B band surface brightness vs. absolute magnitude relation and of the Kormendy
relation has originally shown a strong, apparent dichotomy in the behavior of dwarf and giant ellipticals.
While in dwarfs the effective or central surface brightness increases with luminosity, 
the opposite trend is seen in giants (e.g. Ferguson \& Binggeli 1994; Graham \& Guzman 2003).
An opposite trend between giants and dwarfs has been also observed in the Kormendy relation (e.g. Kormendy 1985; Capaccioli et al. 1992).  
This surprising result has been originally interpreted as a clear indication that 
dwarf ellipticals are not the low luminosity extension of giants but rather an independent class of objects.
The recent study of Graham \& Guzman (2003) based on HST data has however shown that this dichotomy
is only apparent since it is due to a gradual steepening of the central radial profile with luminosity: 
the radial light profile of ellipticals is characterized by a Sersic law, with an index $n$ 
increasing from $\sim$ 1 in dwarfs ($M_B$ $\sim$ -13) to $\sim$ 4 in brighter galaxies 
until the detection of core formation ($M_B$ $\sim$ -20.5), this last probably related to the 
presence of a massive black hole (Faber et al. 1997; C\^ot\'e et al. 2006). The different behavior of E and dE in the 
surface brightness vs. absolute magnitude relation and in the Kormendy relation is just due to
a gradual increase of $n$.
Graham \& Guzman (2003) thus concluded that dE appear to be the low luminosity extension 
of massive ellipticals and suggested that {\it "the mechanisms of how dE and E galaxies collapsed to form stars are
therefore expected to be similar"}.\\
Several recent observational evidences and simulations, however, seem to indicate that local group dwarf 
spheroidals (Mayer et al. 2006), Virgo cluster dwarf ellipticals 
(Barazza et al. 2002; Conselice et al. 2003; van Zee et al. 2004; 
Mastropietro et al. 2005; Lisker et
al. 2006a,b; 2007, 2008, Lisker \& Han 2008; Michielsen et al. 2008) or generally dwarf spheroidals in other
clusters such as Coma (Smith et al. 2008), Perseus (Penny \& Conselice 2008) or in the SDSS
(Haines et al. 2007) might be late-type galaxies recently perturbed by an hostile environment
through ram-pressure stripping or galaxy harassment. 
In a recent work based on a multifrequency analysis of dwarf ellipticals in the Virgo cluster combined with
multizone model of the chemical and photometric evolution of galaxies Boselli et al. (2008) have shown how several observational properties of dE
are consistent with those of star forming discs that recently entered the cluster environment loosing
on a short time scale ($\leq$ 150 Myr) their gas reservoir and thus abruptly stopped their star 
formation activity becoming quiescent systems.
It is thus natural to wonder whether the classical scaling relations such as the surface brightness vs. absolute 
magnitude relation and the Kormendy relation of dwarf ellipticals could be produced by environmental effects 
or are still a clear indication that dE have been formed by the same process than giant ellipticals, that 
the most recent models of hierarchical galaxy formation indicate as major merging (De Lucia et al. 2006). \\
This work is the continuation of what presented in Boselli et al. (2008), where most off the specific 
aspects of the comparison between models and observations have been presented. The surface brightness 
luminosity relation has been already discussed, although in a different form, in Boselli et al. (2008).
Here we consider it in its general form ($<\mu_e(B)>$ vs. $M_B$) and combined with the effective radius 
$R_e(B)$ vs. absolute magnitude relation and the Kormendy relation. The aim of the present paper is 
that of comparing our observations to ram-pressure stripping model predictions and to the dynamical simulations
of harassed galaxies in clusters. In particular we want to investigate whether 
the $<\mu_e(B)>$ vs. $M_B$, $R_e(B)$ vs. $M_B$ and the Kormendy relations can be univocally taken,
as claimed by Graham \& Guzman (2003), as a clear indication that dE have been shaped through the same
formation process than giant ellipticals. The comparison of our model predictions 
with observations indeed indicate that dE might have been formed through a totally different evolutionary process
than their giant counterparts and not results from a hierarchical assembly of matter
as indicated by cosmological models.
 
  \begin{figure*}
   \centering
   \includegraphics[width=12cm,angle=0]{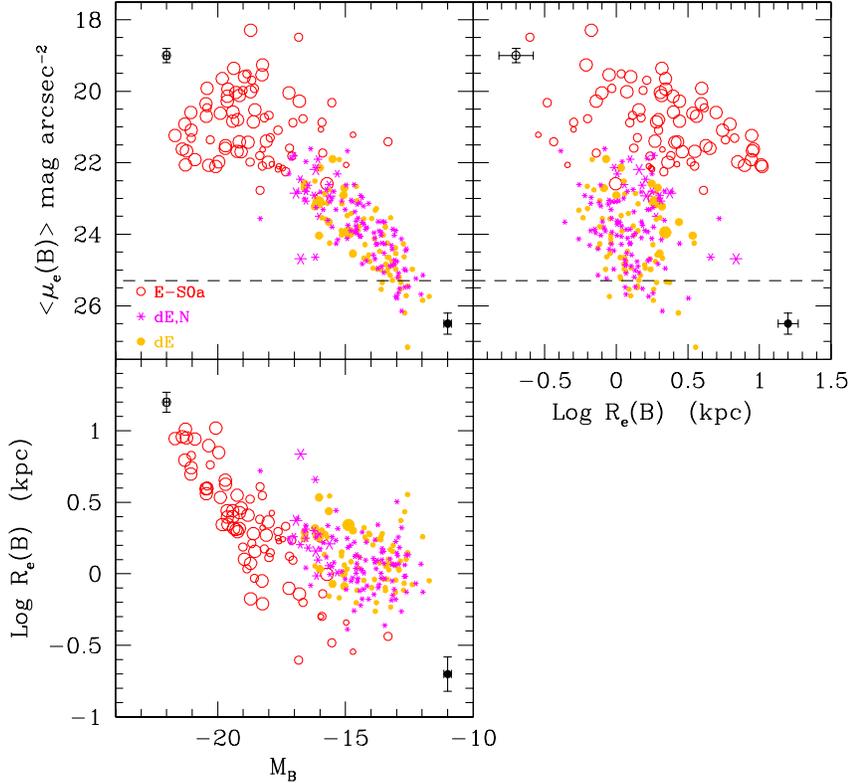}
   \caption{The effective surface
   brightness $<\mu_e(B)>$ (upper left) and radius $R_e(B)$ (lower left) 
   vs. B band absolute magnitude
   relations and the Kormendy relation ($<\mu_e(B)>$ vs. radius $R_e(B)$) (upper right)
   for early-type galaxies: red, open circles are for ellipticals and lenticulars (E-S0-S0a), magenta
   asterisks for nucleated dwarf ellipticals (dE,N), orange, filled dots for dE. 
   Sizes of the symbols have been chosen
   according to the light concentration index: large for $C_{31}(B)$ $>$5; medium
   for 3.5 $<$ $C_{31}(B)$ $\leq$5 and small for $C_{31}(B)$ $\leq$3.5. The horizontal
   dashed line gives the VCC surface brightness detection limit (25.3 mag arcsec$^{-2}$). 
   Typical error bars for giants and dwarfs are given.
    }
   \label{grahamE}
  \end{figure*}

  \begin{figure*}
   \centering
   \includegraphics[width=12cm,angle=0]{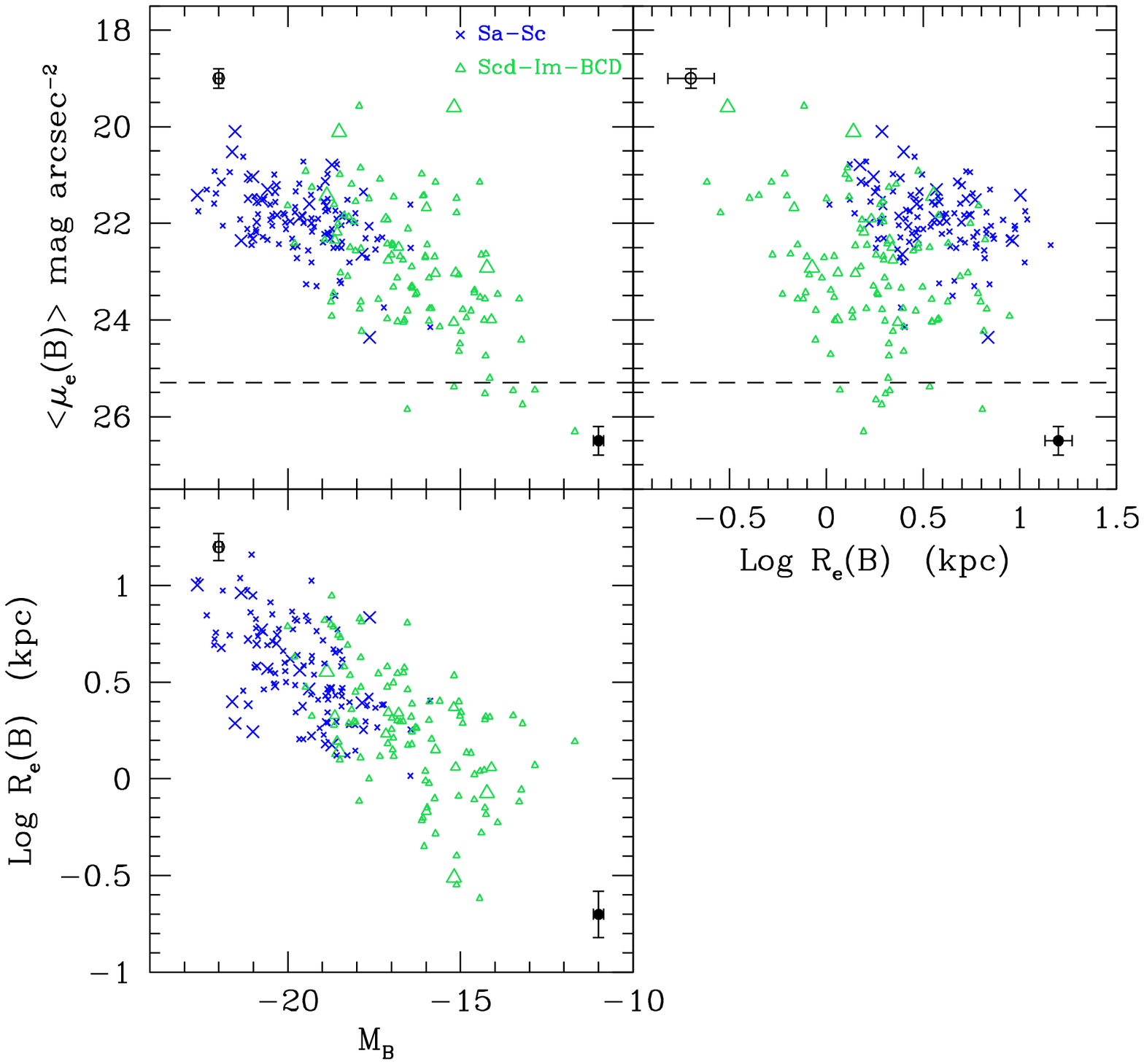}
   \caption{The effective surface
   brightness $<\mu_e(B)>$ (upper left) and radius $R_e(B)$ (lower left) 
   vs. B band absolute magnitude
   relations and the Kormendy relation ($<\mu_e(B)>$ vs. radius $R_e(B)$) (upper right)
   for late-type galaxies (blue crosses for Sa-Sc, green open triangles for Scd-Im-BCD). 
   Sizes of the symbols have been chosen
   according to the light concentration index: large for $C_{31}(B)$ $>$5; medium
   for 3.5 $<$ $C_{31}(B)$ $\leq$5 and small for $C_{31}(B)$ $\leq$3.5.}
   \label{grahamS}
  \end{figure*}
  
\section{The sample}

The present analysis is based on the Virgo Cluster Catalogue
(VCC) of Binggeli et al. (1985) which is an optically selected sample 
complete to $m_B$ $<$18 mag (that for a distance of 17 Mpc gives $M_B$ $\le$ -13.15)
at a limiting surface brightness of $\mu$ = 25.3 mag arcsec$^{-2}$.
Thanks to its proximity, the Virgo cluster is an ideal target to sample 
at the same time both massive and dwarf galaxies without introducing any systematic 
bias in the two galaxy populations do to distance uncertainties \footnote{Unless specified, 
we define as dwarf ellipticals both dE and dwarf spheroidals (dS0).}.
Galaxies analyzed in this work are all bona-fide Virgo
cluster members, whose distances have been assigned following the
subcluster membership criteria of Gavazzi et al$.$ (1999) which is based on 
combined position and redshift data (now available for 83\% of the objects). 
The selected sample 
includes a total of 1017 galaxies, 126 of which are classified as E or S0, 142
as Sa-Sc, 479 as dE or
dS0, 270 as Scd-Sd, Im, BCD, dE/Im or ? in the VCC.

\section{The data}

The B band scaling relations discussed in the present communication have been
reproduced using imaging photometry collected during different observational
campaigns combined with data available in the literature.
The major source of B band imaging data for dwarf ellipticals are 
Gavazzi et al. (2001; 2005), while for massive quiescent and star forming
objects Boselli et al. (2003).\\
Light profile decomposition has been done following the procedures
described in Gavazzi et al. (2000). For each image we extracted the
following structural parameters: the effective radius $r_e$, defined as the radius
including half of the total light, the effective surface brightness $<\mu_e>$,
the mean surface brightness within $r_e$ and the concentration index $C_{31}$,
the ratio of the radii including 75\% and 25\% of the total light.
Typical uncertainties are 0.1 and 0.15 mag in $M_B$, 0.2 and 0.3 mag arcsec$^{-2}$
in $<\mu_e>$, 0.07 and 0.12 in Log $R_e$ and 0.2 and 0.3 in $C_{31}$ for giant and dwarf
galaxies respectively.
We remind that the concentration index parameter is a model independent tracer of 
the shape of the light profile of galaxies. The Sersic profile decomposition, 
probably the most appropriate for reproducing quiescent systems is meaningless for star forming
objects generally characterized by both a bulge and a disc component.
In quiescent system the concentration index is strongly related with the 
Sersic index $n$ ($C_{31}(B)$ = 1.49 $\times$ $n$ + 1.33; Gavazzi et al. 2005), 
while in spirals values of $C_{31}(B)$ $\leq$ 3 are typical of pure exponential profiles.
We thus decided to use the light concentration index parameter $C_{31}(B)$ in order 
to compare in a quantitative way the properties of the light profiles 
of early- and late-type galaxies. \\
Given the heterogeneous nature of the data used in the present work, we
decided to limit our analysis to effective parameters, and in particular to 
the effective surface brightness excluding the generally used
central surface brightness $\mu_0$ whose measure is strongly affected by 
the quality of the seeing (Graham \& Guzman 2003; Gavazzi et al. 2005). 
Furthermore we compare observations to multizone model predictions 
which do not implement specific treatment of the central regions and thus 
do not allow us to obtain an accurate 
estimate of the central surface brightness (see next section).
Images, light profiles and structural properties of galaxies analyzed in this
work can be found on the GOLDMine database (Gavazzi et al. 2003).\\

  \begin{figure*}
   \centering
   \includegraphics[width=16cm,angle=0]{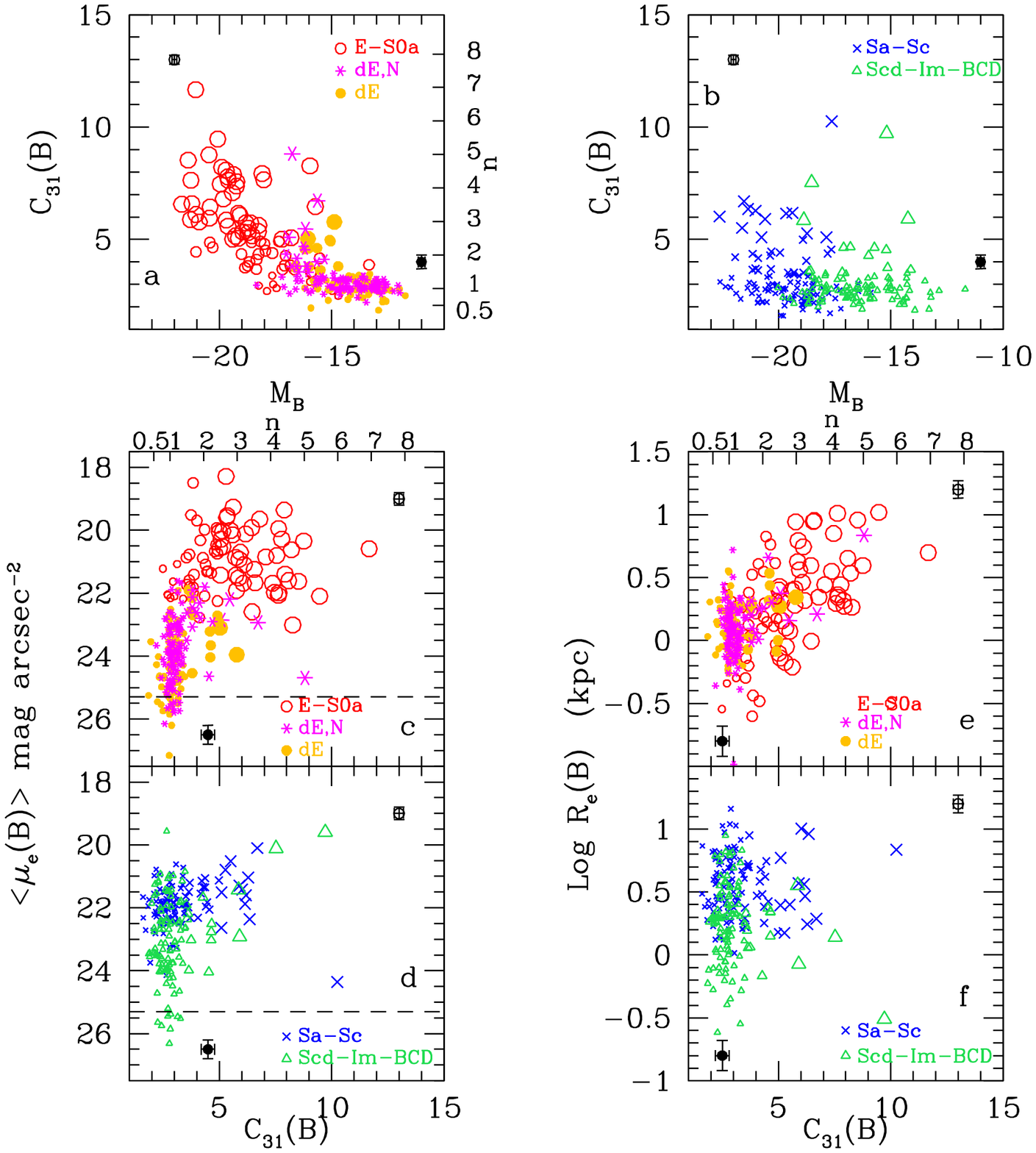}
   \caption{The relationship between 
   the concentration index $C_{31}(B)$ and the absolute magnitude $M_B$ 
   for early-type (a) and for late-type galaxies
   (b), the effective surface brightness $<\mu_e(B)>$ and
   the concentration index $C_{31}(B)$ for early-type (c) and for late-type galaxies
   (d) and the effective radius $R_e(B)$ and
   the concentration index $C_{31}(B)$ for early-type (e) and for late-type galaxies (f).
   Symbols are as in Fig. \ref{grahamE} and \ref{grahamS}. 
   The corresponding Sersic index $n$  valid for early-type galaxies has been determined from the relation 
   $C_{31}(B)$ = 1.49 $\times$ $n$ + 1.33 (Gavazzi et al. 2005).} 
   \label{MBc31_new}
  \end{figure*}

\section{The models}

The observed B band scaling relations of dwarf ellipticals are compared to 
those predicted by two different models of disc galaxies in high density environments:
the first one is the ram-pressure stripping model with subsequent quenching of the
star formation activity extensively described in Boselli et al. (2008), the second one is 
the galaxy harassment model of Mastropietro et al. (2005).\\
The evolution of galaxies in a ram-pressure stripping model is traced using the multi-zone chemical and 
spectro-photometric models of Boissier \& Prantzos (2000), updated with 
an empirically-determined star formation law (Boissier et al$.$ 2003a) 
relating the star formation rate to the total-gas surface densities, 
and modified to simulate a ram-pressure stripping event induced 
by the interaction with the cluster IGM. 
The free parameters in this grid of models are the spin
parameter, $\lambda$ and the rotational velocity, $V_C$.  These
two parameters are theoretical quantities, although $V_C$ should be
similar to the asymptotic value of the rotation curve at large
radii. The spin parameter is a dimensionless measure of the specific angular momentum
(defined in e.g. Mo et al. 1998). Its value in spirals ranges
typically between $\sim$ 0.02 for relatively compact galaxies and has a queue at large values
($\ga$ 0.07) corresponding to low surface brightness galaxies (Boissier et al. 2003b), although
$\sim$ 50\% of the galaxies (including LSB) have values in the range 0.3 $\leq$ 
$\lambda$ $\leq$ 0.6. 
In the models of Boissier \& Prantzos (2000) the total mass varies as $V_C^3$, the scale-length as
$\lambda \times V_C$. Star formation histories depend on the infall
timescales, which are a function of $V_C$, so that 
roughly speaking, $V_C$ controls the stellar mass accumulated during
the history of the galaxy, and $\lambda$ its radial distribution.\\
The ram-pressure exerted by the hot and dense IGM (Gunn \& Gott 1972)
on galaxies crossing the cluster with velocities 
of $\sim$ 1000 km s$^{-1}$ can easily remove the galaxy ISM, in particular the 
atomic hydrogen located outside the optical disc, quenching their
star formation activity (Boselli \& Gavazzi 2006).
The ram-pressure stripping event is here simulated by assuming a gas-loss 
rate inversely proportional to the potential of the galaxy,
with an efficiency depending on the IGM gas density radial profile of the Virgo 
cluster given by Vollmer et al. (2001).\\
For simplicity we will consider here only a ram-pressure stripping model with efficiency
$\epsilon_0$ = 1.2 M$\odot$ kpc$^{-2}$ yr$^{-1}$ (see Boselli et al. 2008 for details).
This efficiency, typical for the gas stripped spiral galaxies in Virgo, can be taken 
as the average value needed to produce the dwarf elliptical galaxy population
through ram-pressure stripping. We also consider only
ram-pressure stripping since it was previously shown that starvation
does not seem to reproduce the observed properties of stripped galaxies (Boselli et al. 2006;
2008).\\
The harassment model is taken from Mastropietro et al. (2005): this is based on
N-body simulations of disc galaxies within a $\Lambda$CDM cluster with 10$^7$ particles,
where the hierarchical growth and galaxy harassment are simulated self-consistently.
These simulations indicate that most of the galaxies undergo major structural modifications 
even at the outskirt of the cluster, with a large fraction of them transforming from late-type
rotating systems into dwarf spheroidal hot systems.
These N-body simulations are well adapted for comparison since they have been defined to 
reproduce the evolution of a relatively low-luminosity late-type galaxy 
(absolute magnitude $M_B$ = -17.08) in a cluster of mass similar to Virgo.
Total masses and effective radii for 13 perturbed simulated galaxies have been kindly provided by C. Mastropietro, and
have been transformed into B band luminosities assuming a constant $M/L_B$ ratio of 6 (and $M/L_B$ = 4.5 
for the unperturbed model), as indicated in Mastropietro et al. (2005).\\

\section{The scaling relations}

The available set of data allows us to reconstruct the effective surface
brightness $<\mu_e(B)>$ and radius $R_e(B)$ vs. B band absolute magnitude
relations and the Kormendy relation ($<\mu_e(B)>$ vs. radius $R_e(B)$)
for both early (E, S0-S0a, dE-dS0) and late-type (Spirals, Im, BCD)
galaxies in the Virgo cluster (respectively in Fig. \ref{grahamE} and \ref{grahamS}).
\noindent  
The surface brightness vs. absolute magnitude relation of quiescent systems shows the well known different
behavior of massive and dwarf galaxies, with the surface brightness increasing with luminosity in
dwarfs and decreasing in giants (e.g. Ferguson \& Binggeli 1994; Graham \& Guzman 2003). 
The effective radius increases with luminosity in giant ellipticals and lenticulars, while
it is almost constant (albeit with a huge scatter) in dwarfs with absolute magnitudes $M_B$ $>$
-18. A substantial difference between giants and dwarfs can be seen in the Kormendy relation, where
a significant dichotomy in the effective surface brightness vs. radius distribution is is present
between  dE and E-S0-S0a.
Considering the strong relationship between the Sersic index $n$ and the concentration 
index $C_{31}(B)$ shown by Gavazzi et al. (2005;  see their Fig.  12), the size variation of
the adopted symbols along the relations is consistent with the apparent dichotomy in the structural properties of
dwarf and giants is due to an increase of the Sersic index $n$ with luminosity (Graham \& Guzman
2003).\\
The surface brightness vs. absolute magnitude relation of star forming systems is similar to that
of dwarf ellipticals, with the surface brightness increasing with luminosity (see Fig.
\ref{grahamS}), although with a flatter slope and a significant larger scatter. The effective
radius increases with absolute magnitude in star forming galaxies for all luminosities but with a
flatter slope than in giant ellipticals. The Kormendy relation of early-type spirals (Sa-Sc) and 
ellipticals and that of late-type spirals and irregulars (Scd-Im-BCD) and dE respectively 
are basically similar, although the difference in between the two quiescent populations seems more
pronounced than for the star forming systems.\\
The relationship between the absolute magnitude, the effective surface 
brightness and radii and the concentration index
are given in Fig. \ref{MBc31_new}.
The large majority of both nucleated and non dwarf ellipticals have concentration indices $\sim$ 3,
thus Sersic indices $n$ $\sim$ 1. All have effective surface brightnesses $\geq$ 21.5 mag
arcsec$^{-2}$ and most of them have effective radii in the range 0.5 $\leq$ $R_e(B)$ $\leq$ 2 kpc.
Giant ellipticals and lenticulars have light concentration indices $C_{31}(B)$ $>$ 3 (thus
Sersic indices $n$ $>$ 1), effective surface brightnesses brighter than 23 mag
arcsec$^{-2}$ and span a larger range in effective radii than dE (0.3 $\leq$ $R_e(B)$ $\leq$ 10 
kpc).  Low luminosity star forming and quiescent galaxies ($M_B$ $\geq$ -18) all have small ($C_{31}(B)$ $\sim$ 3) 
light concentration indices. The light concentration index of massive galaxies 
is systematically higher in ellipticals and lenticulars ($C_{31}(B)$ $\geq$ 5 for
$M_B$ $\leq$ -20) than in spirals ($C_{31}(B)$ $\leq$ 6). 
Late-type galaxies share the same effective surface brightness and radius vs. $C_{31}(B)$
relationships than dwarf ellipticals (both populations are dominated by roughly exponential
profiles), although star forming systems have on average slightly higher surface brightnesses and
larger radii than dwarf ellipticals.\\
These scaling relations are consistent with those observed in other clusters: we notice however that 
quiescent dwarfs in Virgo have slightly larger (but consistent) mean effective radii with a significantly
more dispersed distribution than Antlia (Smith Castelli et al. 2008) and Coma (Graham \& Guzman 2003) dwarfs, being
$R_e(B)$ = 1.41 $\pm$ 0.79 kpc for all the dE of the sample, $R_e(B)$ = 1.40 $\pm$ 0.87 kpc
for the nucleated and $R_e(B)$ = 1.43 $\pm$ 0.66 kpc for the non nucleated ones. This difference might be
due to large statistical uncertainties in the Antlia (36 galaxies) and Coma (18 galaxies) samples
with respect to Virgo (187 objects) or to a slightly different sampled range in luminosity 
and surface brightness.\\

To test possible effects related to the galaxy position within Virgo we compared the 
$<\mu_e(B)>$ vs. $M_B$, $R_e(B)$ vs. $M_B$ and the Kormendy relations for
galaxies located within and outside half the cluster virial radius (2.82$^o$, 0.84 Mpc).
No significant differences have been observed between galaxies located close to the core
or at the periphery of the cluster.

\section{Comparison with model predictions}

Ram-pressure models show how different
structural parameters change once discs stop their star formation activity
because of gas removal. Model predictions are compared to observations in Fig. 
\ref{grahamall}. Since models represent disc galaxies without bulges, they cannot be used to
predict the variation of the light concentration index which 
is a bulge sensitive parameter. We thus first compare model predictions
to the scaling relation determined for galaxies with roughly exponential profiles ($C_{31}(B)$
$\leq$ 3.5, small symbols in Fig. \ref{grahamall}).

\begin{figure}
   \centering
   \includegraphics[width=12cm,angle=0]{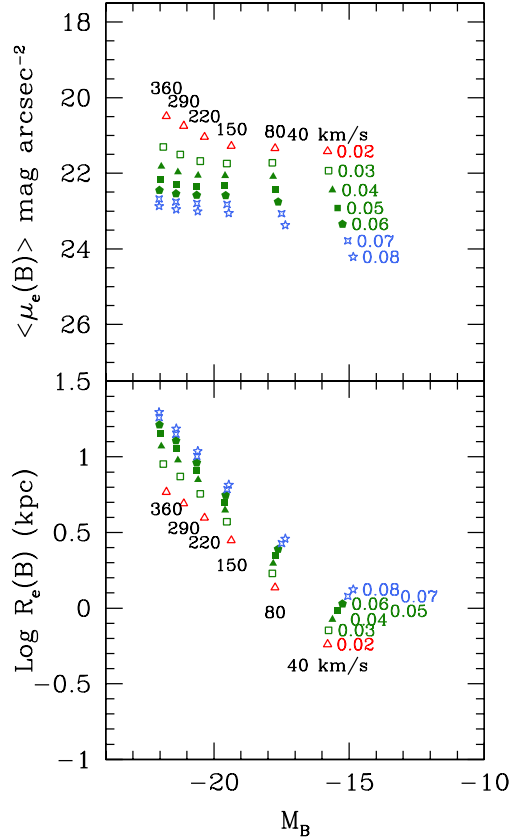}
   \caption{The model B band effective surface brightness (upper panel) and radius (lower panel) as a function of the
   absolute magnitude for different spin parameters (from 0.02 to 0.08) and different velocities (from 40 to 360 km
   s$^{-1}$) in unperturbed, disc galaxies. This indicates the scatter expected in various relationships due to the
   distribution of the spin parameter.
   }
   \label{spin}
  \end{figure}
\begin{figure*}
\centering
\includegraphics[width=12cm,angle=0]{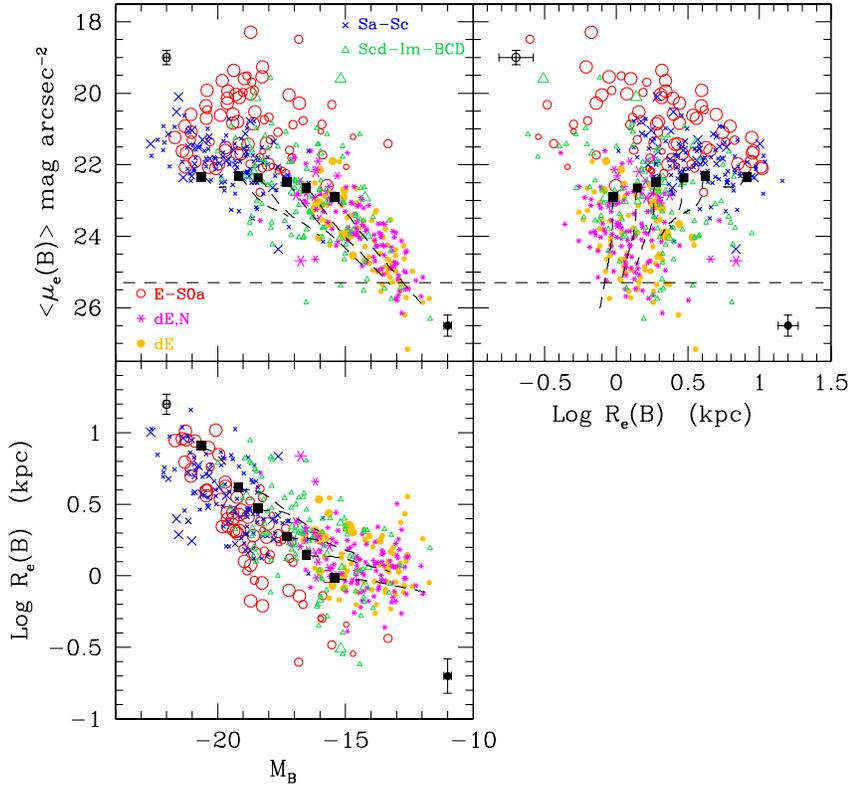}
\caption{Comparison of the model predictions to the scaling relations for all sample galaxies. 
Symbols are as in Fig. \ref{grahamE} and \ref{grahamS}.
Model predictions for unperturbed galaxies of spin parameter
$\lambda$ = 0.05 and $V_C$ = 40, 55, 70, 100, 130 and 220 km s$^{-1}$
are indicated with filled black squares. Model
predictions for a ram-pressure stripping event at different epochs are
indicated by black dashed line.}
\label{grahamall}
\end{figure*}

The observed effective surface brightness vs. absolute magnitude relation (upper left panel in Fig.
\ref{grahamall}) shows a weak decrease of the
effective surface brightness with decreasing luminosity in bright objects (mostly spirals), and
a steeper relation at lower luminosities (dwarf ellipticals). The models indicate that this trend
can be reproduced by stopping the star formation activity of star forming discs. While the
perturbations in massive galaxies ($V_C$ = 220 km s$^{-1}$) are minor, in low luminosity objects 
($V_C$ $\leq$ 70 km s$^{-1}$) the surface brightness and the absolute magnitude are 
significantly reduced with respect to that of a similar unperturbed object. \\
Models also predict the observed variation of the effective radius with absolute magnitude, with 
a decrease of $R_e(B)$ with decreasing luminosity down to $M_B$ $\sim$ -17 (spirals) and a roughly constant 
value at lower luminosities, as indeed observed in dwarf ellipticals. 
Model predictions indicate that the Kormendy relation 
($<\mu_e(B)>$ vs. radius $R_e(B)$) of dwarf ellipticals can result from the transformation of star forming galaxies 
once their activity is stopped because of gas removal after a ram-pressure
stripping event. We remind that a large scatter in the effective surface brightness ($\sim$ 1.5 mag) and radius (0.4
in dex) are predicted by models if we consider that the spin parameter is expected to vary in normal galaxies 
in the range between 0.03 $\leq$ $\lambda$ $\leq$ 0.06 (see Fig. \ref{spin}). More extreme values of $\lambda$,
although possible, are expected to be rare. The observed scatter in the $R_e(B)$ 
and $<\mu_e(B)>$ parameters in Fig. \ref{grahamall} can thus be ascribed to the distribution in
the spin parameter expected in normal, disc galaxies.
Among spiral galaxies, the most discrepant from models of unperturbed objects are,
as expected, those with prominent bulges ($C_{31}(B)$ $>$ 3.5, large and medium blue crosses and green triangles), 
this component not being considered by our models.  \\

\begin{figure*}
\centering
\includegraphics[width=12cm,angle=0]{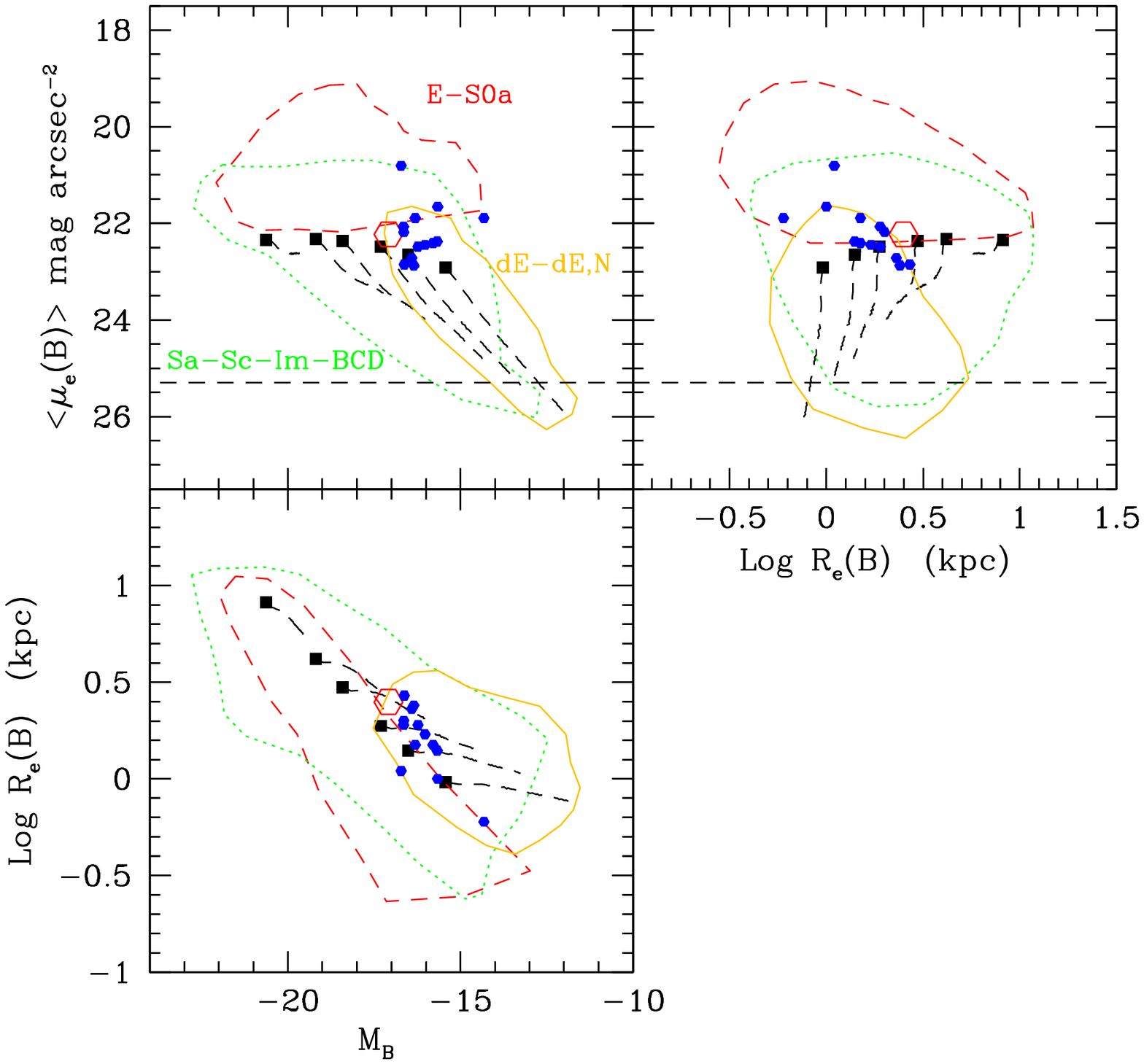}
\caption{Comparison of our ram-pressure stripping model predictions 
and those predicted by Mastropietro et al. (2005) for galaxy harassment 
to the scaling relations for all sample galaxies. 
Our ram-pressure model predictions for unperturbed galaxies of spin parameter
$\lambda$ = 0.05 and $V_C$ = 40, 55, 70, 100, 130 and 220 km s$^{-1}$
are indicated with filled black squares, and their evolution after a
ram-pressure stripping event by black dashed lines. The unperturbed galaxy 
for the models of Mastropietro et al. (2005) is indicated by a large, red hexagon while
the simulated harassed galaxies by small, blue filled hexagons. Continuum, dotted and
dashed contours indicate the distribution of dE-dE,N (orange), spiral, Im and BCD (green), 
ellipticals and lenticulars (red) galaxies as determined by eye from our own observations.}
\label{mastropietro2}
\end{figure*}

Figure \ref{mastropietro2} shows the effects of ram-pressure stripping (filled black squares 
indicate unperturbed galaxies of different rotational velocity, black dashed lines the evolution
after a ram-pressure stripping event) and galaxy harassment (the red empty pentagon shows the unperturbed
galaxy, blue filled pentagons the 13 different perturbed models) with respect
to the observations, here indicated by continuum, dotted and
dashed contours for the distribution of dE-dE,N (orange), spiral, Im and BCD (green), 
ellipticals and lenticulars (red) respectively.
The comparison of the observed scaling relations with the dynamical simulations of Mastropietro et al. (2005)
indicate that harassed galaxies are more compact and have comparable or higher surface 
brightness than unperturbed objects (see Fig. \ref{mastropietro2}), probably
because of the formation of bars (Mastropietro et al. 2005).
Their absolute magnitude decreases after the interaction because of mass loss.
Dynamical interactions have thus significantly different effects than those induced by a ram-pressure stripping event.\\
Harassed galaxies have structural properties in between those of low-luminosity and
and massive ellipticals. It is however difficult to state whether harassed galaxies fails in reproducing
the low-luminosity, low-surface brightness dwarf ellipticals at the origin of the 
$<\mu_e(B)>$ vs. $M_B$, $R_e(B)$ vs. $M_B$ and the Kormendy relations
just because dynamical simulations are only available for an intermediate
mass disc galaxy.

The structural properties of giant ellipticals and lenticulars (red open circles) are significantly
different from those of massive spirals (blue crosses) and, as expected, cannot be produced by
a simple quenching of the star formation activity of spirals after a ram-pressure stripping event.
The dynamical simulations of Mastropietro et al. (2005), available only for a relatively low-mass
late-type galaxy, prevent us to see whether the enhanced surface brightness of early-type galaxies,
and in particular of S0s, with respect to that of late-types of similar luminosity,
could result from galaxy harassment. To answer to this important question it would be necessary
to extend the simulation of Mastropietro et al. (2005) in order to cover the dynamic range in stellar
mass of the late-type galaxy population of the Virgo cluster.\\
We can thus conclude that most of the dwarf ellipticals
have structural properties consistent with being ram-pressure gas stripped, low luminosity
star forming galaxies, as stated in Boselli et al. (2008).
The lack of any observed trend in the scaling relations with the clustercentric distance is in agreement
with this view since the gas stripping phenomenon and the subsequent stopping of the star formation
activity acts on relatively short times scales ($\leq$ 150 Myr, Boselli et al. 2008)
if compared to the cluster crossing time, which in Virgo is of the order of 1.7 Gyr. It is thus
almost impossible to observe objects in their Im/BCD - dE transitionary phase. 
The observed morphology-segregation effect in dwarfs (Ferguson \& Binggeli 1994) 
is indeed consistent with this picture.

\section{Discussion and conclusion}

The analysis presented so far is totally consistent with the idea that
the origin of the different behavior of dwarf and massive early-type galaxies 
along the B band effective surface brightness and effective radius vs. absolute
magnitude relations and along the Kormendy relation often described in the literature 
is deeply related to the shape of their radial light
profiles, as suggested by Graham \& Guzman (2003). We indeed observe that the effective surface brightness
decreases (and the effective radius increases) with luminosity only in early-type galaxies with high 
concentration indices or high Sersic indices $n$,
while it increases (it is almost constant) whenever $C_{31}(B)$ $<$ 5 or $n$ $<$ 2.5.
Massive ($M_B$ $\leq$ -20) ellipticals and lenticulars all have $C_{31}(B)$ $\geq$ 5, 
while quiescent dwarfs ($M_B$ $\geq$ -18) have $C_{31}(B)$ $\sim$ 3 consistent with a 
roughly exponential profile.\\
In late-type galaxies the effective surface brightness and the effective radius
monotonically increase with luminosity, but with flatter slopes and larger scatters than in
low luminosity and massive ellipticals respectively. They all
have $C_{31}(B)$ $\sim$ $\leq$ 6, with $C_{31}(B)$ $\sim$ 3 in $\sim$ 70\% of the massive objects. \\ 
The comparison with models have shown that the distribution of quiescent galaxies along the aforementioned scaling relations
cannot be univocally taken as a proof of a similar formation process for dwarf and giants, 
as stated by Graham \& Guzman (2003). The lack of dynamical simulations
for the whole range in luminosity of the disc galaxy population
prevents us to say whether the scaling relations observed in E-dE galaxies 
here discussed can be the result of the harassment suffered by spirals entered into the cluster.\\
The observed relations between the effective surface brightness, 
effective radius and absolute magnitude of low luminosity ellipticals are
consistent with their formation after a recent and rapid decrease of the star formation activity
of low luminosity spirals recently entered into the cluster medium and swept of their
gas by a ram-pressure stripping event, as claimed by Boselli et al. (2008).
This evolutionary path is thus significantly different than that followed by massive
ellipticals for which hierarchical models of galaxy formation predict they
formed most of their stars at early epochs ($z$ $\geq$ 2) and later
assembled through major merging events (De Lucia et al 2006). A different evolution 
of massive and dwarf ellipticals is also confirmed by the analysis of their
spectro-photometric properties and of their [$\alpha$/Fe] ratios, both indicating a rapid
formation of the bulk of the stellar population in the early universe in massive ellipticals
while a rather continuous and moderate star formation activity for dwarf systems 
(Thomas et al. 2005; Renzini 2006).\\
As a general conclusion we can say that the present work is a further indication that
the interpretation of single scaling relations in terms of evolution is not always 
straightforward but rather needs to be included in a complete and coherent study
combining multifrequency observations with model predictions.

\begin{acknowledgements}

This research has made use of the NASA/IPAC Extragalactic Database (NED) which is operated by the 
Jet Propulsion Laboratory, California Institute of Technology, under contract with the National 
Aeronautics and Space Administration, and of the GOLD Mine database. We are grateful to C. Mastropietro
who provided us with the results of their N-body dynamical simulations, and the anonymous referee
for precious comments and suggestions which helped improving the quality of the manuscript.

\end{acknowledgements}


\begin{thebibliography}{}

\bibitem[]{}Barazza, F., Binggeli, B., Jerjen, H., 2002, A\&A, 391, 823

\bibitem[]{}Bender, R., Burstein, D., Faber, S., 1993, ApJ, 411, 153

\bibitem[]{}Binggeli B., Sandage A., Tammann G., 1985, AJ, 90, 1681

\bibitem[]{}Boissier, S. \& Prantzos, N., 2000, MNRAS, 312, 398

\bibitem[]{}Boissier, S., Prantzos, N., Boselli, A. \& Gavazzi, G., 2003a, MNRAS, 346, 1215 

\bibitem[]{}Boissier, S., Monnier Ragaigne, D., Prantzos, N., van Driel, W., Balkowski, C. \& O'Neil, K., 2003b, MNRAS,
343, 653 
    

\bibitem[]{}Boselli, A. \& Gavazzi, G., 2006, PASP, 118, 517




\bibitem[]{}Boselli, A., Gavazzi, G., Donas, J., Scodeggio, M., 2001, AJ, 121, 753

\bibitem[]{}Boselli A., Lequeux J., Gavazzi G., 2002, A\&A, 384, 33


\bibitem[]{}Boselli A., Gavazzi G., Sanvito G., 2003, A\&A, 402, 37



\bibitem[]{}Boselli A., Boissier, S., Cortese, L., Gil de Paz, A., Seibert, M., Madore, B., Buat, V., Martin, C.,
2006, ApJ, 651, 811

\bibitem[]{}Boselli A., Boissier, S., Cortese, L., Gavazzi G., 2008, ApJ, in press (Feb 20)

\bibitem[]{}Bower, R.G., Lucey, J.R., Ellis, R.S., 1992, MNRAS, 254, 601

\bibitem[]{}Capaccioli, M., Caon, N., D'Onofrio, M., 1992, MNRAS, 259, 323

\bibitem[]{}Conselice, C., O'Neil, K., Gallagher, J., Wyse, R., 2003, ApJ, 591, 167


\bibitem[]{}C\^ot\'e, P., Piatek, S., Ferrarese, L., et al., 2006, ApJS, 165, 57


\bibitem[]{}De Lucia, G., Springel, V., White, S., Croton, D., Kauffmann, G., 2006, MNRAS, 366, 499

\bibitem[]{}Djorgovski, S., Davis, M., 1987, ApJ, 313, 59

\bibitem[]{}Dressler, A., Lynden-Bell, D., Burstein, D., Davies, R., Faber, S., Terlevich, R., Wegner, G., 1987, ApJ,
313, 42

\bibitem[]{}Faber, S., Tremaine, S., Ajhar, E., et al., 1997, AJ, 114, 1771

\bibitem[]{}Ferguson, H., Binggeli, B., 1994, A\&ARev, 6, 67



\bibitem[]{}Gavazzi, G., Pierini, D. \& Boselli, A., 1996, A\&A, 312, 397


\bibitem[]{}Gavazzi, G., Boselli, A., Scodeggio, M., Pierini, D., Belsole, E., 1999, MNRAS, 304, 595

\bibitem[]{}Gavazzi, G., Franzetti, P., Scodeggio, M., Boselli, A., Pierini, D., 2000, A\&A, 361, 863


\bibitem[]{}Gavazzi, G., Zibetti, S., Boselli, A., Franzetti, P., Scodeggio, M., Martocchi, S., 2001, A\&A, 372, 29



\bibitem[]{}Gavazzi, G., Boselli, A., Donati, A., Franzetti, P. \& Scodeggio, M., 2003, A\&A, 400, 451


\bibitem[]{}Gavazzi, G., Donati, A., Cuccati, O., Sabatini, S., Boselli, A., Davies, J., Zibetti, S., 2005, A\&A,
430, 411




\bibitem[]{}Graham, A., Guzman, R., 2003, AJ, 125, 2936

\bibitem[]{}Gunn, J.E., \& Gott, J.R.I., 1972, ApJ, 176, 1

\bibitem[]{}Haines, C., Gargiuolo, A., La Barbera, F., Mercurio, A., Merluzzi, P., Busarello, G., 2007, MNRAS, 381, 7

\bibitem[]{}Haines, C., Gargiuolo, A., Merluzzi, P., 2008, MNRAS, 385, 1201



\bibitem[]{}Kormendy, J., 1985, ApJ, 432, L63

\bibitem[]{}Kormendy, J., \& Kennicutt, R., 2004, ARA\&A, 42, 603

\bibitem[]{}Lisker, T., Grebel, E., Binggeli, B., 2006a, AJ, 132, 497

\bibitem[]{}Lisker, T., Glatt, K., Westera, P., Grebel, E., 2006b, AJ, 132, 2432

\bibitem[]{}Lisker, T., Grebel, E., Binggeli, B., Glatt, K., 2007, ApJ, 660, 1186

\bibitem[]{}Lisker, T., Grebel, E., Binggeli, B., 2008, AJ, 135, 380

\bibitem[]{}Lisker, T., Han, Z., 2008, ApJ, in press

\bibitem[]{}Mayer, L., Mastropietro, C., Wadsley, J., Stadel, J., Moore, B., 2006, MNRAS, 369, 1021

\bibitem[]{}Mastropietro, C., Moore, B., Mayer, L., Debattista, V., Piffaretti, R., Stadel, J.,
2005, MNRAS, 364, 607

\bibitem[]{}Michielsen, D., Boselli, A., Conselice, C., et al., 2008, MNRAS, 385, 1374

\bibitem[]{}Mo, H.~J., Mao, S., \&  White, S.~D.~M.\ 1998, MNRAS, 295, 319


\bibitem[]{}Penny, S., Conselice, C., 2008, MNRAS, 383, 247

\bibitem[]{}Renzini, A., 2006, ARA\&A, 44, 141



\bibitem[]{}Smith, R., Marzke, R., Hornschemeier, A., et al., 2008, MNRAS, in press

\bibitem[]{}Smith Castelli, A., Bassino, L., Richtler, T., Cellone, S., Aruta, C., Infante, L., 2008, MNRAS, in press


\bibitem[]{}Thomas, D., Maraston, C., Bender, R., Mendes de Oliveira, C., 2005, ApJ, 621, 673

\bibitem[]{}Tully, B., Fisher, J., 1977, A\&A, 54, 661

\bibitem[]{}Tully, B., Mould, J., Aaronson, M., 1982, ApJ, 257, 527

\bibitem[]{}Van Zee, L., Skillman, E., Haynes, M., 2004, AJ, 128, 121

\bibitem[]{}Visvanathan, N., Sandage, A., 1977, ApJ, 216, 214

\bibitem[]{}Vollmer, B., Cayatte, V., Balkowski, C. \& Duschl, W., 2001, ApJ, 561, 708

\bibitem[]{}Zaritsky, D., Kennicutt, R., Huchra, J., 1994, ApJ, 420, 87



\end{thebibliography}
\end{document}